\documentclass[fleqn,usenatbib]{mnras}

\usepackage{newtxtext,newtxmath}
\usepackage{mathptmx}

\usepackage[T1]{fontenc}

\DeclareRobustCommand{\VAN}[3]{#2}
\let\VANthebibliography\thebibliography
\def\thebibliography{\DeclareRobustCommand{\VAN}[3]{##3}\VANthebibliography}

\usepackage{graphicx}	
\usepackage{amsmath} 

\usepackage[switch]{lineno}

\usepackage{graphicx,pstricks}
\usepackage{color}

\usepackage{url}
\usepackage{xspace}

\newcommand{\gls}{\texttt{PS~J2305+3714}\xspace}
\newcommand{\WHT}{\texttt{WHT/ISIS}\xspace}
\newcommand{\HST}{\texttt{HST}\xspace}

\title[Double lensed quasar PS~J2305+3714]{Time-delay and lens galaxy redshift in the doubly imaged quasar PS~J2305+3714}

\author[O. A. Burkhonov et al.]{
    O. A. Burkhonov$^{1}$\thanks{E-mail: boa@astrin.uz},
    V. N. Shalyapin$^{2,3}$,
    A. V. Sergeyev$^{4,5,6}$,
	Sh.E. Nurmamatov$^{7,1}$,
\newauthor
	Sh. A. Ehgamberdiev$^{1,8}$,
    T.A. Akhunov$^{8,1}$,
    F. Dux,
    F. Courbin$^{9, 10, 11}$,
    M.M. Muminov$^{12}$
\\
$^{1}$Ulugh Beg Astronomical Institute, 33 Astronomicheskaya St., Tashkent 100052, Uzbekistan\\
$^{2}$Departamento de F\'\i sica Moderna, Universidad de Cantabria,
    Avda. de Los Castros s/n, E-39005 Santander, Spain\\
$^{3}$O.Ya. Usikov Institute for Radiophysics and Electronics, National
    Academy of Sciences of Ukraine, 12 Acad. Proscury St., UA-61085
    Kharkiv, Ukraine\\
$^{4}$Institute of Astronomy, Kharkiv V.N. Karazin National University, 35 	Svobody Sq., UA-61022 Kharkiv, Ukraine\\
$^{5}$Observatoire de la C\^ote d'Azur, Boulevard de l'Observatoire, 06304 Nice, France\\
$^{6}$Institute of Radio Astronomy, National Academy of Sciences of Ukraine, 4 Mystetstv St., UA-61002 Kharkiv, Ukraine\\
$^{7}$Chirchik State Pedagogical University. 104, Amir Temur str., Chirchik city, Uzbekistan\\
$^{8}$National University of Uzbekistan, Tashkent, Uzbekistan\\
$^{9}$ICC-UB Institut de Ciéncies del Cosmos, University of Barcelona, Martí Franqués, 1, E-08028 Barcelona, Spain\\
$^{10}$ Institució Catalana de Recerca i Estudis Avançats (ICREA), Passeig de Lluís Companys 23, 08010 Barcelona, Spain\\
$^{11}$Institut d'Estudis Espacials de Catalunya (IEEC), 08860 Castelldefels, Barcelona, Spain\\
$^{12}$Andijan State University, 129 Universitet st., Andijan city 170100, Uzbekistan \\
}

\date{Accepted XXX. Received YYY; in original form ZZZ}

\pubyear{\the\year{}}

\begin{document}
\label{firstpage}
\pagerange{\pageref{firstpage}--\pageref{lastpage}}
\maketitle


\begin{abstract}

We present results from a seven-season (2018-2024) monitoring campaign of the gravitationally lensed quasar system \gls using the 1.5-m Maidanak Telescope in the optical R-band. From these data, the amplitude of possible microlensing variability does not exceed 10 mmag on a 7-yr timescale. Additionally, we measure a time delay of $t_{AB}$ = 52.2$\pm$2.5 days, with the brighter image A leading. Long-slit \WHT spectroscopy refines the quasar redshift to $z_s$ = 1.791 and provides the first measurement of the lens redshift, $z_d$ = 0.473. The flux ratios of the quasar images in the MgII $\lambda$2800 emission line and in the adjacent continuum are nearly identical, indicating minimal microlensing effects in the spectral domain, which is consistent with the very weak microlensing signal in the time domain. Using precise astrometry from recent HST imaging and the MgII flux ratio, we also build two simple mass models for the lens system. The close agreement between the measured delay and those predicted by the mass models, measured redshifts, and a concordance $\Lambda$CDM cosmology, confirms the robustness of our results and highlights \gls as a promising system for future time-delay cosmography.

\end{abstract}

\begin{keywords}
gravitational lensing: strong -- quasars: individual: PS J2305+3714 -- galaxies: redshifts -- cosmology: observations
\end{keywords}



\section{Introduction}
\label{sec:introd}
Gravitational lensed quasars can be used for the Hubble constant measuring. This idea was suggested a long ago \citep{Refsdal1964} but only in the last decade it is transformed into a powerful astrophysical tool known as "Time Delay Cosmography" \citep{Birrer2024} that is able to concurrent with other available methods and complement them. The characteristic feature of this method is the possibility to obtain useful constraints even from the analysis of a single target; however, for a more robust evaluation, a collection of well-studied objects is necessary. Such a set of 6 lensed quasars was used, e.g., in \cite{Wong2020}, which led to a 2.4\% precision measurement in accordance with the local value of the Hubble constant. Certainly, the estimated value depends on details of the specific study and on the involved sample size, so the extension of a useful lensed quasar set is an actual problem.

Few hundred galaxy-scale gravitationally lensed quasars have already been confirmed, and forecast studies predict that upcoming wide-field synoptic surveys will expand the sample to thousands \citep{Oguri2010,Collett2015}.
The Vera C.\ Rubin Observatory's Legacy Survey of Space and Time (LSST) alone is expected to deliver $\sim$8\,000 lenses, while \textit{Euclid} and the Nancy Grace Roman Space Telescope will add thousands more, especially at higher redshift.

Only a minority of these systems are suitable for precision ``time-delay cosmography'' because they must exhibit clean, multiply-imaged variability and permit accurate mass-model inference \citep{Gilman2020}. 
For each useful lens, the relative delays between quasar images need to be measured to better than about $5\%$ in order to keep the cosmological error budget competitive. The Strong-Lens Time-Delay Challenge showed that once this threshold is met, lens-model systematics—not timing errors—dominate the final uncertainty \citep{Liao2015,Dobler2015}.

Longer intrinsic delays (tens to hundreds of days) are favoured, because they mitigate the impact of seasonal gaps in ground-based data \citep{Tewes2013}.  Achieving a few percent precision typically requires densely-sampled, multi-season photometric monitoring \citep{Courbin2011}.  
The COSMOGRAIL programme has pioneered this strategy, yielding benchmark sub-percent delays for lenses such as HE\,0435$-$1223 and RX\,J1131$-$1231 \citep{Bonvin2017, Millon2020, Dux2025}.  Combining these precise delays with state-of-the-art lens modelling, the H0LiCOW and TDCOSMO collaborations have already measured the Hubble constant to $2$-$3\%$ precision \citep{TDCOSMO2025}.

Monitoring observations of gravitationally lensed quasars at the Maidanak Observatory \citep{Ehgamberdiev2018} began in 1995 \citep{Vakulik1997}. For several of these systems, time delays were measured using Maidanak data \citep{Tsvetkova2010, Bekov2024}, and multicolor photometric monitoring was conducted for some targets \citep{Akhunov2017, Goicoechea2020}.

In this paper, we provide above mentioned essential information required for the Hubble constant evaluation for another lensed quasar system \gls. This system was discovered by \citep{Lemon2018} from \emph{Gaia} observations. The authors presented astrometry based on Pan-STARRS \emph{grizY} images, developed a mass model, and extracted the quasar source redshift from the William Herschel Telescope (\WHT) long-slit spectra. 
Later \cite{Shajib2021} observed \gls with Keck/NIRC2 infrared camera, refined astrometry, and NIR photometry of the lens galaxy and quasar images and presented their mass model, which predicts the time delay between the quasar components as $58.00^{\text{+0.15}}_{-0.18}$ days calculated for fiducial redshifts of lens galaxy $z_d$=0.5 and source $z_s$=2.0.

We present the results of our long-term optical monitoring of the system, including the construction of light curves and the measurement of the time delay between the quasar images in Section~\ref{sec:monitoring}. In Section~\ref{sec:spec}, we reprocess the archive \WHT spectral observations to determine the redshift of the lensing galaxy. Section~\ref{sec:hst} focuses on the analysis of available \HST imaging data, providing accurate astrometric parameters of the system. The observational results are then used to construct simple lens mass models and to compare the corresponding theoretical time delays with the measured value in Section~\ref{sec:MassModel}. The conclusions and discussion are provided in the final Section~\ref{sec:concl}.

\section{Maidanak optical monitoring}
\label{sec:monitoring}
\subsection{Observational data}
\label{sec:ccd}
The target was monitored using the 1.5-m Maidanak Telescope with a Bessel $R$ filter over seven observational seasons from 2018 to 2024. During six of those seasons (2018-2022 and 2024), observations were conducted with the SNUCAM camera \citep{Im2010}, a $4\mathrm{k} \times 4\mathrm{k}$ CCD with a pixel scale of 0.269\arcsec/pixel, a gain of 1.45~$e^-$/ADU, and a readout noise of 1.45~$e^-$. In the 2023 season, the Andor XL camera was used instead, featuring a similar $4096 \times 4108$ format, a pixel scale of 0.267\arcsec/pixel, a gain of 1.55~$e^-$/ADU, and a higher readout noise of 2.61~$e^-$. Each observing night typically included either five 300-second exposures or eight 180-second exposures. The full dataset comprises 1472 frames, with a measured median seeing of $1.29\arcsec \pm 0.13\arcsec$.

The standard reduction for both CCD cameras includes bias subtraction and flat-fielding correction. The Andor XL camera was operated at a higher temperature -80$^{\circ}$ C (vs -108$^{\circ}$ C in SNUCAM), and dark current subtraction was required.

\begin{figure}
	\centering
	\includegraphics[width=9cm]{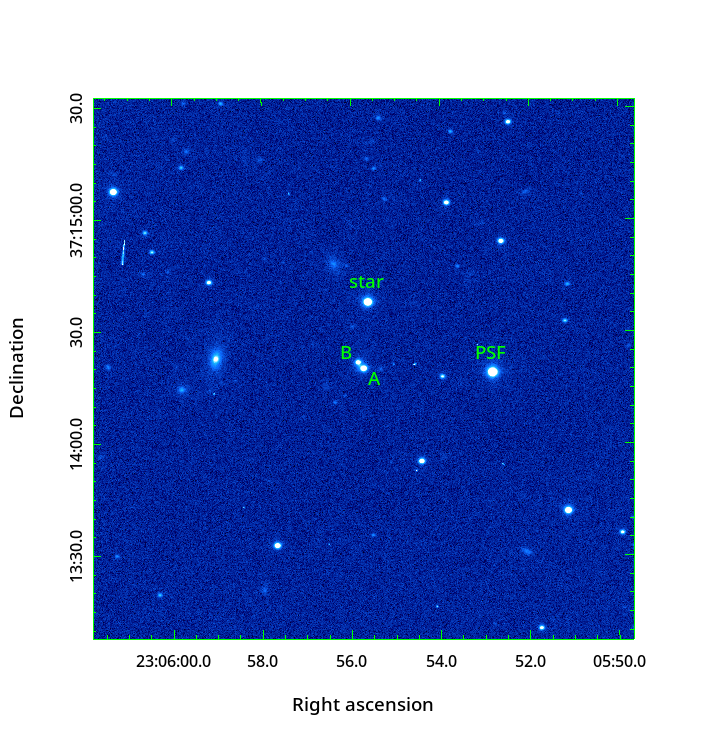}
	\caption{The central region of a typical 300-s exposure. This particular frame was taken on the night of 2018-07-06 under 0.96$\arcsec$ seeing. The two quasar images are labeled A and B.}
	\label{fig:FOV}
\end{figure}

The central region of the full field of view around the lensed quasar is shown in Fig.~\ref{fig:FOV}.

\subsection{Light curves}
\label{sec:lcs}

The angular separation between the quasar images is only $\sim$2.2\arcsec, requiring point-spread function (PSF) photometry to accurately separate the fluxes of the two components and the lensing galaxy. For this purpose, we employ the \texttt{IMFITFITS} software \citep{McLeod1998}, which has been widely used in the analysis of gravitationally lensed systems (e.g., \citealt{Gil-Merino2018}). The model includes two point sources representing the quasar images and an extended component describing the lensing galaxy. The galaxy light distribution is modeled using a de Vaucouleurs profile. The relative positions of the two point sources and the galaxy center, as well as the structural parameters of the galaxy—effective radius, ellipticity, and major-axis position angle—are fixed to values derived from \HST imaging (see Section~\ref{sec:hst}).

The model components are convolved with a PSF constructed from a nearby reference star (labeled “PSF” in Fig.~\ref{fig:FOV}). The software then computes the optimal fluxes for the two quasar images, the lens galaxy, and the background level by minimizing the residuals between the observed and model light distributions. In the first iteration, the fluxes of both quasar components and the galaxy are treated as free parameters. In the second iteration, the ratio of the galaxy's flux to the PSF star's flux is fixed to its average value obtained in the first step, and only the quasar fluxes, background level and the position of A quasar image are re-optimized.

The \texttt{IMFITFITS} procedure was applied to each individual frame in the dataset. Of the total 1472 frames, only 53 (approximately 3\%) yielded unsatisfactory photometric results and were excluded from further analysis. The remaining measurements were averaged nightly to construct the final light curves.

Photometric uncertainties were estimated by comparing magnitude differences between adjacent epochs separated by no more than 2.5 days. The resulting mean photometric errors are $0.005$~mag for component~A and $0.009$~mag for component~B. These mean uncertainties were then scaled for each night according to the inverse of the estimated signal-to-noise ratio.

\begin{figure}
\centering
\includegraphics[width=9cm]{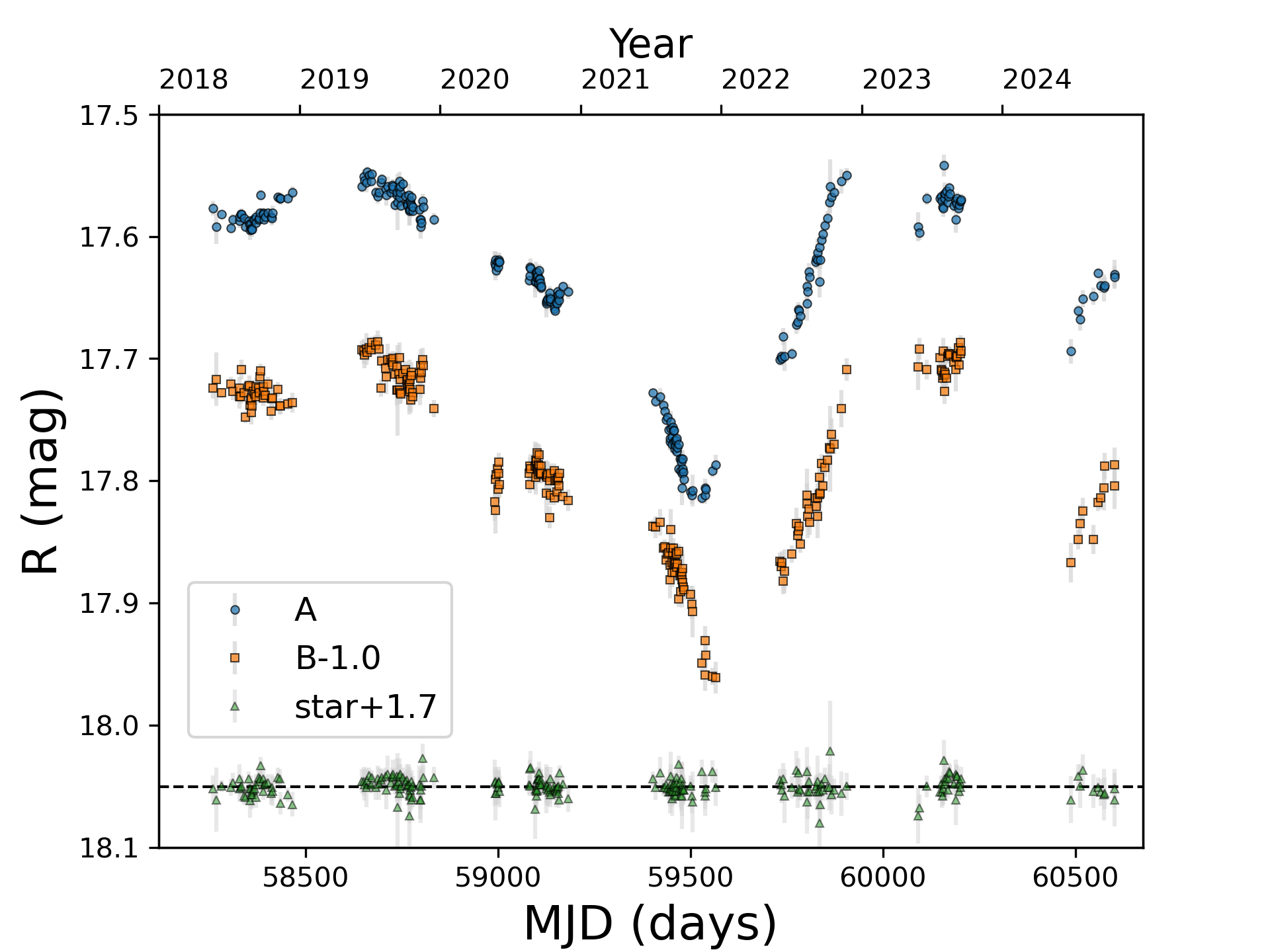}
\caption{The light curves of quasar components A, B, and a comparison star. The measured magnitudes of image B and the comparison star are shifted by -1.0 mag and +1.7 mag, respectively, to facilitate comparison with image A.}
\label{fig:lcs}
\end{figure}

To transform relative magnitudes to absolute values, we adopted the $R$-band magnitude of the PSF reference star from the USNO-B1.0 catalog, listed as 15.61~mag. The calibrated magnitudes of the quasar components, along with their estimated uncertainties, as well as the magnitudes and errors of the comparison star (labeled “star” in Fig.~\ref{fig:FOV}), are available via the CDS. The corresponding light curves are also shown in Fig.~\ref{fig:lcs}.

\subsection{Time delay}
\label{sec:timedelay}
At present, a variety of methods have been developed for estimating time delays between two light curves. While these methods may show differing levels of convergence in numerical experiments with synthetic data \citep{Dobler2015}, they tend to produce consistent results when applied to real observational data. In this study, we employ the dispersion minimization technique \citep{Pelt1996}, which has been widely tested on multiple lens systems and has demonstrated robust performance. Specifically, we adopt a modified $D^2_4$ version of the dispersion method that incorporates a Gaussian weighting function. Compared to traditional rectangular or triangular weighting schemes, the Gaussian function introduces additional smoothness to the results, particularly under variations of its width parameter \citep{Shalyapin2025}. 

\begin{figure}
\centering
\includegraphics[width=9cm]{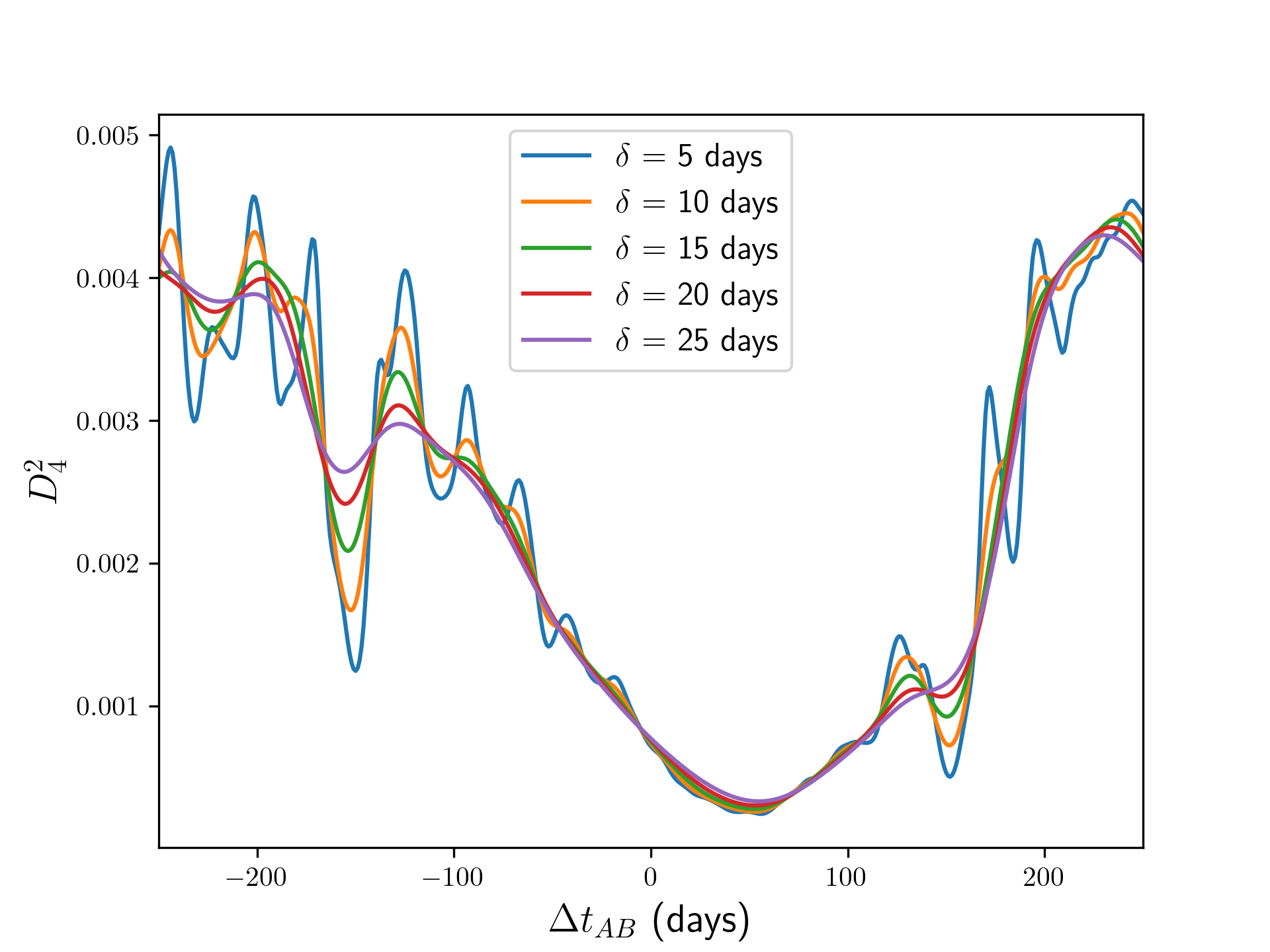}
\caption{The dependency of dispersion $D^2_4$ on trial time delays $\Delta t_{AB}$ for a range of decorrelation lengths $\delta$.}
\label{fig:disper}
\end{figure}

In its simplest form, the dispersion method assumes a constant magnitude offset between two light curves, without accounting for any potential microlensing variability. The dependence of the dispersion function $D^2_4$ on the trial time delay between the quasar components, $\Delta t_{AB}$, is shown in Fig.~\ref{fig:disper}. Since the exact shape of the dispersion curve is sensitive to the width of the weighting function—referred to as the decorrelation length $\delta$, the following results are presented for a range of $\delta$ values from 5 to 25 days. Short decorrelation lengths tend to introduce greater oscillations in the dispersion curve, while excessively long values can lead to systematic biases.

For all tested $\delta$ values, the dispersion function exhibits a well-defined global minimum and increases gradually for larger positive or negative trial delays. Local secondary minima near $\pm$half a year arise due to seasonal gaps in the observational data. Although the precise value of the delay corresponding to the minimum of $D^2_4$ depends somewhat on the choice of $\delta$, this dependence is relatively weak. For $\delta = 10$, 15, and 20 days, the corresponding best-fit time delays are $\Delta t_{AB} = 52.7$, 52.3, and 53.1 days, respectively, with an intrinsic scatter of only 0.4 days.

A more robust estimate of the time delay can be obtained using a bootstrap resampling approach, in which synthetic realizations of the observed light curves are generated by perturbing the measured magnitudes with random Gaussian noise \citep{Pelt1996}. We created 1000 such realizations, where each simulated light curve pair was constructed by adding noise consistent with the estimated photometric uncertainties at each epoch.

\begin{figure}
	\centering
	\includegraphics[width=9cm]{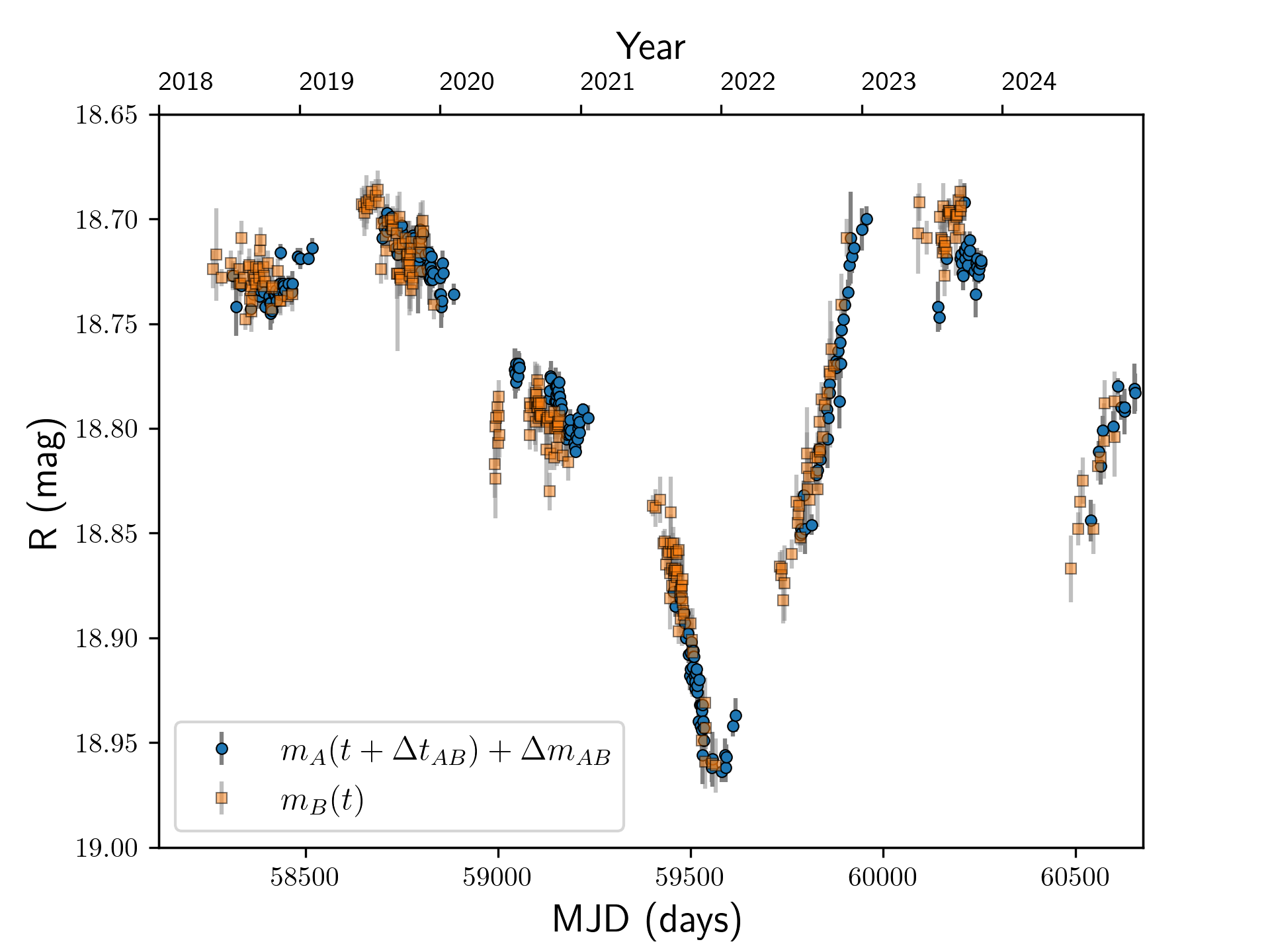}
	\caption{Time delayed and magnitude shifted A light curve and superimposed original B light curve.}
	\label{fig:lcsshifted}
\end{figure}

For each simulated pair, the time delay was calculated using the dispersion method with a fixed decorrelation length of $\delta = 15$ days. The resulting distribution of time delays is unimodal and yields a final estimate of $\Delta t_{AB} = 52.4 \pm 2.3$ days. The corresponding magnitude offsets between the light curves of components A and B are distributed around $\Delta m_{AB} = 1.150 \pm 0.001$~mag. Fig.~\ref{fig:lcsshifted} illustrates the alignment of the light curves, showing component~B alongside the time- and magnitude-shifted light curve of component~A.

Fig.~\ref{fig:lcsshifted} shows that a simple constant magnitude shift aligns the two light curves reasonably well. This indicates that potential temporal microlensing variations during the monitoring period were minimal. To further assess the influence of microlensing, we applied the dispersion method with an incorporated polynomial microlensing model \citep{Shalyapin2025}. Introducing a linear microlensing trend slightly modifies the optimal time delay to $\Delta t_{AB} = 51.8^{\text{+2.2}}_{-2.1}$ days and adjusts the magnitude shift to $\Delta m_{AB} = 1.147 \pm 0.001$~mag.

The fitted linear trend results in a magnitude change of $-0.014 \pm 0.003$~mag between the start and end of the observational campaign. This variation is approximately two orders of magnitude smaller than the central magnitude offset $\Delta m_{AB}$ and remains within the combined observational uncertainties of components~A and B. These results confirm that microlensing effects were negligible over the duration of the monitoring period.

As the final estimate of the time delay, we combine the two cases discussed above, yielding $\Delta t_{AB} = 52.2 \pm 2.5$~days. It should be noted that the applied bootstrap method likely overestimates the confidence intervals to some extent. Ideally, simulated noise should be added to the true, noise-free magnitudes; however, in practice, the bootstrap procedure perturbs the already noisy observed data, effectively adding simulated noise on top of real observational noise.
Nevertheless, we retain the final calculated error to provide a conservative estimate of the time-delay uncertainty. We also applied the spline-based time-delay estimation technique of \citet{Millon2020}, finding that the central value of the delay remains unchanged, although its uncertainty increases to 4 days. 
We note, however, that this broader error interval arises from a method that models both the intrinsic and microlensing variability with splines; the simultaneous reconstruction of both signals may therefore lead to an overestimation of the delay uncertainty.

\section{\WHT spectroscopy}
\label{sec:spec}
The discovery paper by \citet{Lemon2018} presents \WHT spectra of the two quasar components. The strong similarity between these spectra supports the gravitationally lensed nature of the \gls system. While the quasar redshift was estimated as $z_s = 1.78$, no information was provided about the redshift of the lensing galaxy. In this work, we address that gap.

We retrieved the original \WHT observations of two exposures of 600~s each, taken on 2017-09-12 from the public ING data archive, along with the corresponding calibration files. Observations from both the blue (R300B) and red (R158) arms were reduced using standard procedures within the NOIRLab IRAF v2.18 environment \citep{Fitzpatrick2024}. The reduction steps included bias subtraction, overscan correction, trimming, flat-fielding, cosmic-ray rejection, wavelength calibration using Cu, Ne, and Ar arc lamps, background subtraction, and sensitivity correction using the spectrophotometric standard star BD+$28^\circ~4211$. The final result consists of 2D spectra with wavelength dispersion and spatial axes.

The original spectral extraction in the discovery work \citep{Lemon2018} was performed using Gaussian apertures with a width of 0.5\arcsec\ along the spatial axis. In contrast, we adopted a more advanced technique based on spatial profile modeling, following the methods described by \citet{Sluse2007} and \citet{Goicoechea2019}. Specifically, the spatial 1D profile at each wavelength was modeled as a sum of three Moffat functions representing components A, B, and the lensing galaxy G, all sharing a common width. The relative separations between the profile centers were fixed using the astrometric constraints from \HST imaging (Section~\ref{sec:hst}).

During the first iteration, we estimated the position of component~A along the dispersion axis, the Moffat profile width, and the fluxes of components A, B, and G at each wavelength. These estimates were then smoothed with low-order polynomials to define the trace of component~A and the profile width, which were held fixed in a second iteration. In this final step, only the fluxes of A, B, and G were optimized across the dispersion axis.

\begin{figure}
	\centering
	\includegraphics[width=9cm]{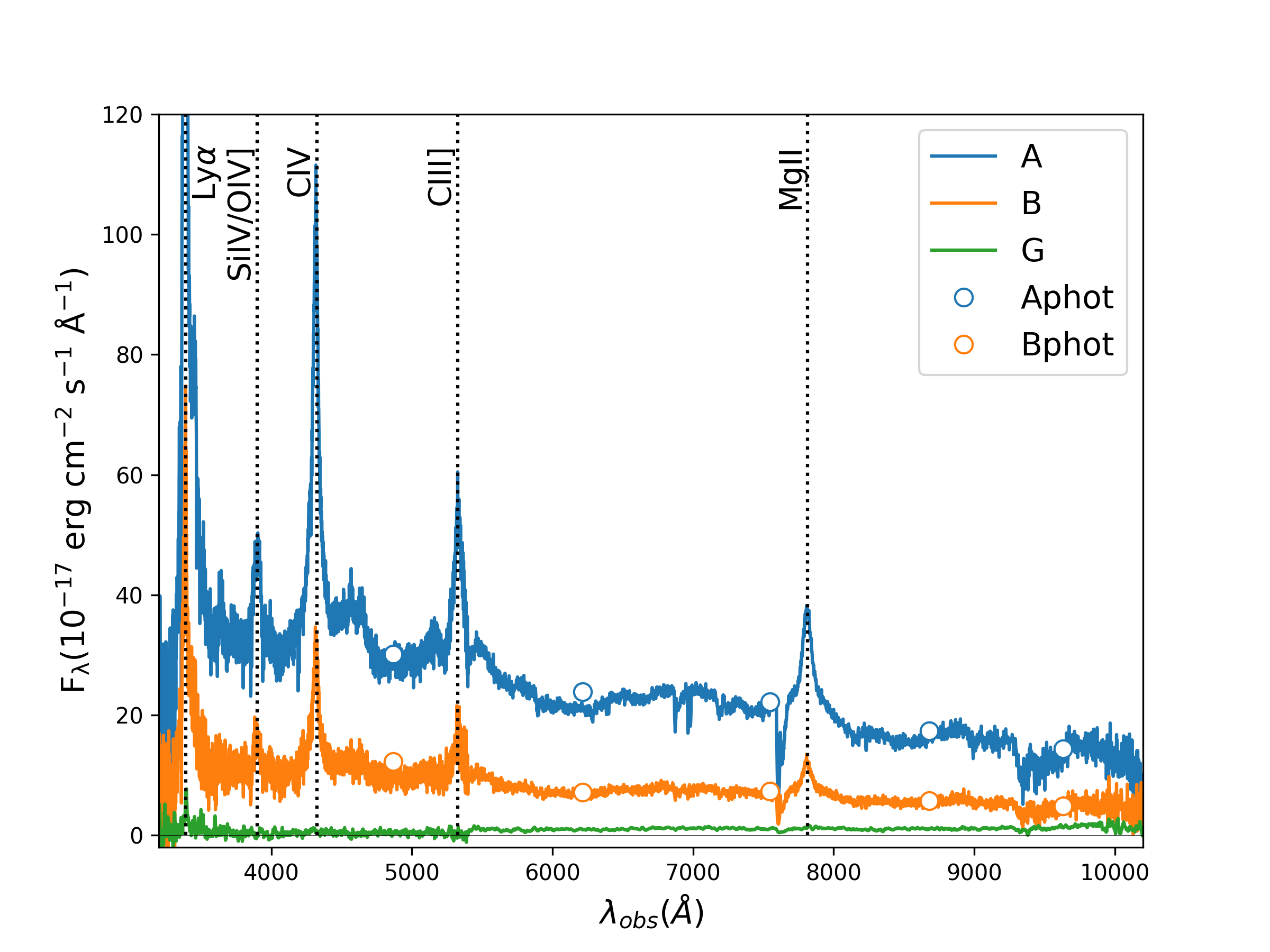}
	\caption{WHT spectra of the quasar components A and B, as well as the lensing galaxy G, observed on 2017-09-12. Open circles indicate Pan-STARRS broadband photometric fluxes for comparison. Several prominent quasar emission lines are labeled.}
	\label{fig:specABG}
\end{figure}

The extracted spectra of both quasar components and the lensing galaxy are presented in Fig.~\ref{fig:specABG}. The quasar spectra closely match those published earlier by \citet{Lemon2018}. For comparison, we also show broadband \textbf{\textit{grizY}} photometric fluxes from Pan-STARRS, obtained several years prior. The good agreement between the \WHT and Pan-STARRS fluxes suggests only moderate variability over the $\sim$3-year interval between the two datasets.

\begin{figure}
	\centering
	\includegraphics[width=9cm]{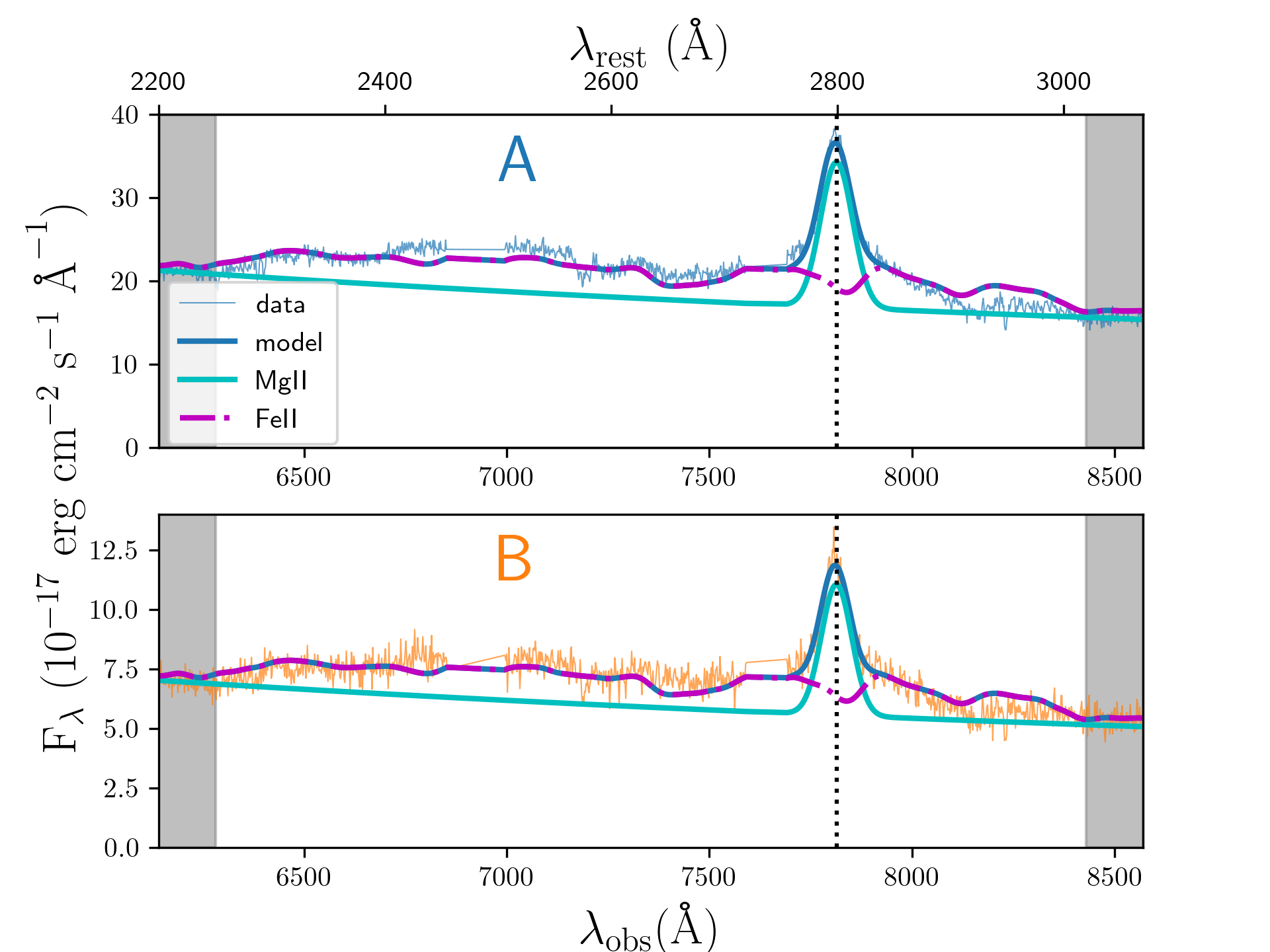}
	\caption{Modeling of the \ion{Mg}{II} $\lambda2800$ emission region in the \WHT\ spectra of components A and B. Absorption features were masked during the fitting procedure.}
\label{fig:specABMgII}

\end{figure}

The extracted \WHT\ spectra provide valuable information in three key aspects. The first two are related to the analysis of the \ion{Mg}{II} $\lambda2800$ emission line, as shown in Fig.~\ref{fig:specABMgII}. The spectral region around \ion{Mg}{II} was modeled using a combination of a power-law continuum, an Fe\,\textsc{ii} pseudo-continuum template \citep{Tsuzuki2006}, and a Gaussian profile for the emission line itself. This analysis yields a refined redshift of the quasar, $z_s = 1.791$, slightly higher than the previously published estimate of $z_s = 1.78$.

In addition, we compared the flux ratios of components B and A in both the \ion{Mg}{II} emission line and the underlying continuum at 2800\,\AA. The measured ratios are B/A(Mg\,\textsc{ii}) = $0.295\pm0.005$ and B/A(cont$_{2800}$) = $0.328\pm0.002$, which are in close agreement. Since the continuum emission originates from the compact accretion disk region and is therefore susceptible to microlensing magnification, while the \ion{Mg}{II} emission arises from a much larger region with negligible microlensing impact, the small difference between the two ratios suggests that microlensing effects are minor. This result supports the conclusion drawn from the light curve analysis in Section~\ref{sec:timedelay}, where no significant microlensing variability was detected. (The flux ratio between two emission lines B/A(Mg\,\textsc{ii}) and underlying continuum B/A(cont$_{2800}$) calculated from a single spectrum do not take into account the internal variability of the quasar at two epochs separated by a time delay, so we do not interpret the moderate difference between B/A(Mg\,\textsc{ii}) and B/A(cont$_{2800}$) as a detection of microlensing.)

\begin{figure}
	\centering
	\includegraphics[width=9cm]{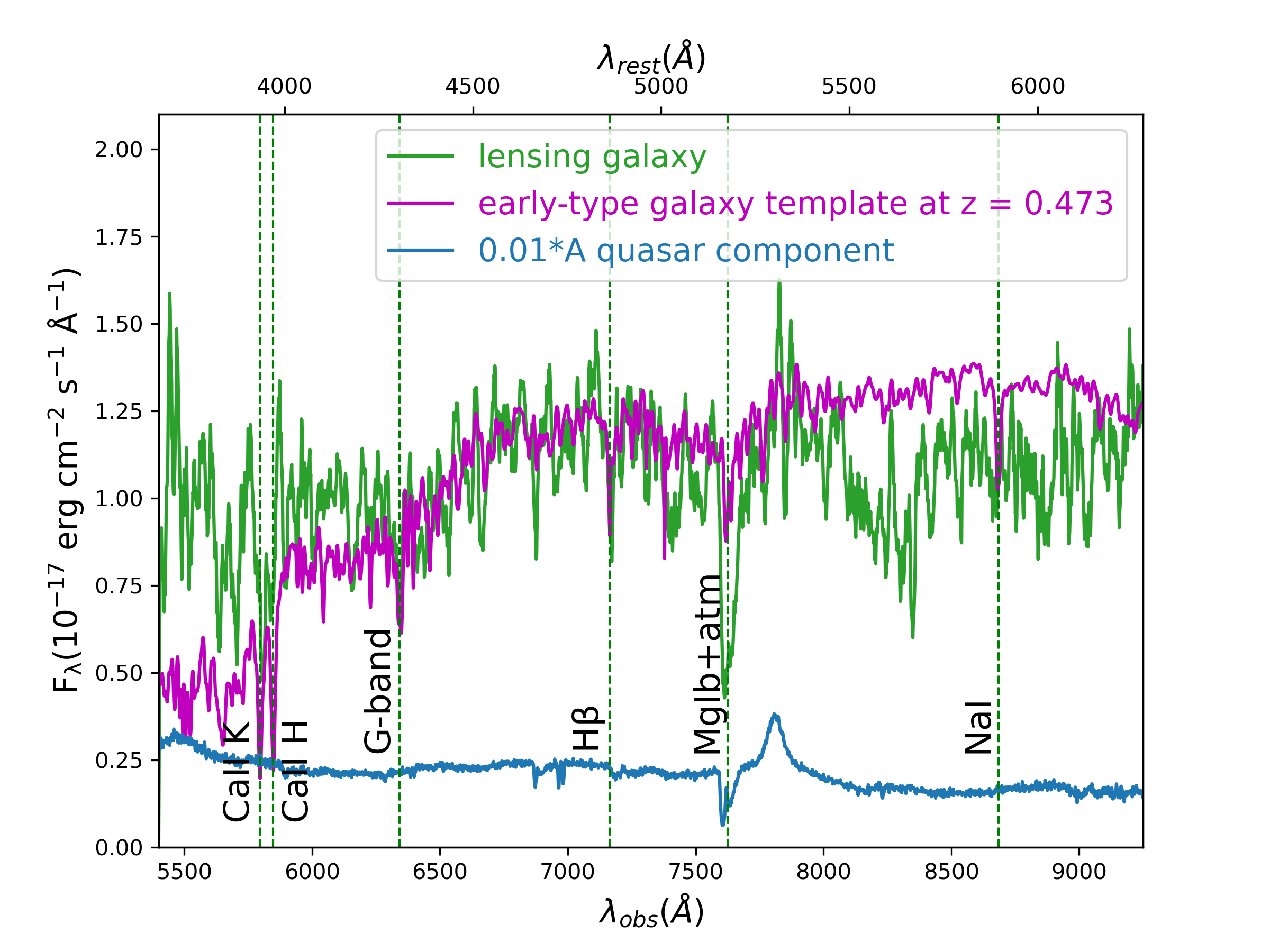}
	\caption{Extracted spectrum of the lensing galaxy (green) compared with an early-type galaxy template (magenta), redshifted to the estimated value of $z_d = 0.473$. The scaled quasar component A (blue) is shown for comparison.}
	\label{fig:specG}
\end{figure}

The comparison in Fig.~\ref{fig:specABG} clearly illustrates that both quasar components, A and B, are significantly brighter than the lensing galaxy G. As a result, the overlapping spatial profiles of the two quasar images introduce substantial contamination to the galaxy spectrum, leading to a relatively noisy extraction. Despite this, the extracted spectrum of the lensing galaxy (Fig.~\ref{fig:specG}) clearly shows prominent absorption features, including the CaII H\&K doublet, G-band, and $H_{\beta}$ which are consistent with those of an early-type galaxy at a redshift of $z_d = 0.473 \pm 0.001$. This newly determined lens redshift, together with the refined quasar redshift, represents an essential input for future determinations of the Hubble constant based on this gravitational lens system.

\section{\HST astrometry}
\label{sec:hst}

Accurate astrometry is critical for constructing precise mass models of gravitational lens systems. Previous studies developed lens models using Pan-STARRS astrometry \citep{Lemon2018} and Keck adaptive optics imaging \citep{Shajib2021}. In this work, we take advantage of recently obtained \HST\ imaging data to improve the astrometric precision.

The \gls\ system was observed with the \HST\ ACS/WFC instrument on 2023-10-14\footnote{HST proposal 17308, PI: Cameron Lemon}. 
Two dithered exposures of 337~seconds each were acquired using the F814W filter. We retrieved these images from the \HST/MAST archive and removed cosmic rays using the
\texttt{Astro-SCRAPPY} \citep{astroscrappy2025} package, 
which is based on the L.A.Cosmic algorithm developed by \citet{vanDokkum2001}.

The separation of the two quasar components from the lensing galaxy requires accurate PSF photometry. The ACS/WFC camera exhibits a strong dependence of the PSF shape on both the telescope focus level and the position on the detector. To account for these variations, a model PSF was generated using the ePSF Webtool \citep{Anand2023} and interpolated to the precise location of the lensed system on the detector.

\begin{figure}
	\centering
	\includegraphics[width=9cm]{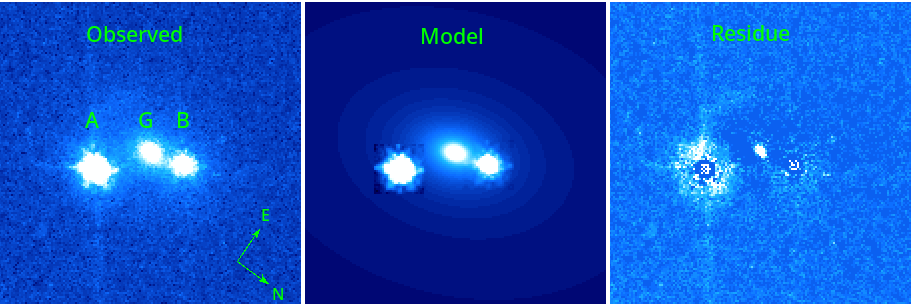}
	\caption{GALFIT modeling of the \gls\ system based on the first \HST/ACS F814W frame. 
	\textit{Left:} Observed image showing the two quasar components (A and B) and the lensing galaxy (G). Component A is visibly saturated. 
	\textit{Middle:} Best-fit model constructed using two PSF-convolved point sources for the quasar images and a de Vaucouleurs profile for the lensing galaxy. 
	\textit{Right:} Residual image after subtracting the model from the observation. Significant residual flux remains near component A due to saturation, and additional structure in the central region of the galaxy may suggest deviations from a pure de Vaucouleurs profile or the presence of arc-like features.}

	\label{fig:galfit3}
\end{figure}

The observed light distribution was modeled as a combination of two point sources representing the quasar components and an extended de Vaucouleurs profile for the lensing galaxy. The positions of the quasar images and the galaxy center, along with the structural parameters of the galaxy, were fitted using the \texttt{GALFIT} software \citep{Peng2002}. Fig.~\ref{fig:galfit3} presents the observed image, the best-fit model, and the residuals for the first ACS/WFC frame; the second exposure yields a nearly identical result. The derived coordinates of the quasar components and the galaxy center are summarized in Table~\ref{tab:astrometry}. The formal differences between the estimated coordinates in the two ACS/WFC frames are small. However, for the subsequent mass modeling, we adopt more conservative positional uncertainties: 0.001\arcsec\ for component~B and 0.003\arcsec\ for the lensing galaxy. Magnitude errors estimated by GALFIT are $\sim$0.02 mag.

\setcounter{table}{1}
\begin{table}[h!]
	\begin{center}
		\caption{Coordinates and magnitudes of the quasar components and the lensing galaxy in the \gls\ system, extracted from ACS/WFC images using \texttt{GALFIT}.}
		\label{tab:astrometry}
		\begin{tabular}{ccccc}
			\hline \hline
			Component & $\Delta$RA ($\arcsec$) & $\Delta$Dec ($\arcsec$) & F814 (mag) \\
			\hline 
			A 		  &  0.000         & 0.000           & 17.38       \\
			B  		  &  1.460         & 1.645           & 18.54       \\
			G         &  1.183         & 0.853           & 18.13       \\
			\hline
		\end{tabular}
	\end{center} 
\end{table}

The elliptical galaxy light distribution is parameterized by an effective radius of R$_{eff}= 2.98\arcsec \pm 0.07\arcsec$, an axial ratio of q$_L=0.69\pm0.01$ and a positional angle of PA$_L=18.6^\circ\pm0.8^\circ$. It is also worth noting the residual light visible in the right panel of Fig.~\ref{fig:galfit3}. The prominent residual near the position of component~A is expected, as this component is saturated in the \HST\ images. Additional residuals in the central region of the lensing galaxy may indicate structural complexity beyond a simple de Vaucouleurs profile. In particular, arc-like features near component~A may be present. These structures, however, are not analyzed further in the current study.

\section{Mass Model}
\label{sec:MassModel}

Previous studies by \citet{Lemon2018} and \citet{Shajib2021} presented mass models of the system based on earlier observational data. In this work, we construct an updated and more refined lens model using the newly determined redshifts, the improved astrometric positions of the quasar components and lensing galaxy, and the \ion{Mg}{II} flux ratio as a proxy for macrolens magnification. The modeling was performed using the \texttt{Gravlens/Lensmodel} software package \citep{Keeton2001a}.

Our mass modelling relies only on observational constraints on the relative positions of the quasar images with respect to the lensing galaxy and the image flux ratio. This small set of constraints is sufficient to achieve an exact fit, with $\chi^2 = 0$, of simple mass models. We explore two widely used models: the Singular Isothermal Sphere with external shear (SIS+shear) and the Singular Isothermal Ellipsoid (SIE) \citep{Keeton2001b}. Both models yield similar results, with predicted time delays\footnote{using the measured redshifts and a $\Lambda $CDM cosmology with $H_0=70\rm \ km \ s^{-1}\ Mpc^{-1}$, $\Omega_m = 0.3$, $\Omega_{\Lambda} = 0.7$} of $\Delta t_{AB} = 54.3$~days (SIS+shear) and $56.9$~days (SIE), compared to the measured value of $52.2 \pm 2.5$~days (see Section~\ref{sec:timedelay}). The corresponding Einstein radii are 1.202\arcsec\ and 1.169\arcsec, and the total magnifications are 5.94 and 5.12, respectively. The axis ratio q$_m = 0.72$ and the position angle PA$_m = 10^\circ$ of the SIE mass distribution closely resemble the corresponding parameters of the light distribution presented in the previous section. The reasonable agreement between the
light and SIE mass structure (q and PA parameters), and between the measured and predicted time delay, seem to indicate that the actual mass distribution is close to isothermal and that external effects (convergence and shear) do not play a relevant role.

As we adopt SIS/SIE profiles with a fixed isothermal slope, our models are only indicative of the actual behaviour. In principle, the mass sheet degeneracy implies that a range of slopes can fit the data. Nevertheless, the agreement between predicted and observed delays is reassuring.

The close agreement between the model-predicted and observed time delays confirms the consistency of the lens configuration and supports the reliability of the measured quantities used in the modeling.

For comparison, we rescale the time delay estimate reported by \citet{Shajib2021}, who found $\Delta t_{AB} = 58.0$ days assuming fiducial redshifts of $z_d = 0.5$ and $z_s = 2.0$, to our measured redshifts of $z_d = 0.473$ and $z_s = 1.791$. The rescaled time delay is $55.2$ days. This estimate was derived using an entirely independent dataset and modeling software, yet it remains in good agreement with our result. Such consistency across different approaches and data sources reinforces the robustness of the time-delay measurement and the lens model.

\section{Conclusions and Discussion}
\label{sec:concl}

The main results of this study can be summarized as follows:
\begin{itemize}
    \item We conducted a 7-year monitoring campaign of the double-lensed quasar system \gls\ during the 2018-2024 period using the 1.5-m Maidanak Telescope in the $R$-band.

    \item From the full dataset, we obtained reliable photometry on 258 nights.

    \item The time delay between the quasar images was measured using the $D^2_4$ dispersion minimization technique \citep{Pelt1996}, yielding $\Delta t_{AB} = 52.2 \pm 2.5$ days. No significant temporal microlensing variability was detected during the monitoring period.
    
    \item Archival spectroscopic data obtained with the 4.2-m William Herschel Telescope were retrieved from the ING archive and reprocessed. We extracted spectra for both quasar components and the lensing galaxy. These spectra enabled us to refine the source redshift and to measure the lens redshift for the first time.. The similarity between the flux ratio in the \ion{Mg}{II} $\lambda2800$ emission lines and the underlying continuum further supports the conclusion that microlensing effects are very weak.
    
    \item Refined astrometry of the system was derived from newly available \HST/ACS imaging in the F814W band.
    
    \item The precise positions of the quasar components and the lensing galaxy, along with the \ion{Mg}{II} flux ratio (used as a proxy for macrolensing magnification) and the updated source and lens redshifts, were used as inputs to simple lens models. Using the \texttt{Gravlens/Lensmodel} software, both the SIS+shear and SIE models produced predicted time delays of 54.3 and 56.9 days, respectively—values that are consistent with our measured delay and with a previously rescaled estimate of 55.2 days reported by \citet{Shajib2021}.
\end{itemize}

The internal consistency of the photometric, spectroscopic, astrometric, and lens-modelling results demonstrates that \gls\ is a robust system, contributing to the growing sample of doubles suitable for time-delay cosmography, with consistent results across multiple datasets and methods. 
In particular, the availability of high-precision measurements of $\Delta t_{AB}$, source and lens redshifts, and image positions establishes a solid foundation for constraining the Hubble constant $H_0$ in upcoming follow-up work using more sophisticated mass modelling and external datasets.

While simple mass models such as SIS+shear and SIE are useful for estimating expected quantities, they cannot fully capture the complexity of the actual mass distribution. At least two important factors have not been accounted for in the present analysis. 
First, the mass density profile of the lensing galaxy may deviate from the isothermal assumption. This deviation could be probed via a measurement of the galaxy's stellar velocity dispersion. Unfortunately, the available \WHT\ spectrum is too noisy for a reliable extraction of this quantity, but the lensing galaxy is sufficiently bright that dedicated observations with an 8-10 meter class telescope could resolve this limitation.
Second, the local environment and line-of-sight structures can introduce external convergence, thereby influencing the inferred mass model and any resulting estimate of $H_0$. This effect can be assessed through integral field spectroscopy or wide-field galaxy surveys around the lens, again requiring deeper observations with large telescopes.

The next steps are well defined: obtaining a reliable measurement of the lensing galaxy's stellar velocity dispersion and characterizing the external convergence along the line of sight are essential. These observational efforts—ideally carried out with large-aperture telescopes that will supply the critical inputs needed for constructing more sophisticated mass models. With these refinements, \gls\ can be transformed into a powerful tool for precision cosmology and an independent probe of the Universe's expansion rate.

\section*{Acknowledgements}

The authors would like to thank their colleagues for helpful discussions and acknowledge the use of the 1.5m Maidanak Telescope and the 4.2-m William Hershel Telescope for the observations presented in this paper. We gratefully acknowledge the financial support provided by the Ministry of Higher Education, Science and Innovation of the Republic of Uzbekistan under project No. IL-5421101855. We thank the anonymous referee for her/his comments, which helped us to improve our manuscript.

\section*{Data Availability}

The extracted light curves and spectra are available via the CDS.
The images and raw data underlying this article are available upon request.



\bibliographystyle{mnras}
\bibliography{MN-25-2635-MJ_rev} 

\bsp	
\label{lastpage}
\end{document}